\documentclass[journal=jacsat,manuscript=article]{achemso}

\usepackage{chemformula} 
\usepackage{soul}
\usepackage[T1]{fontenc} 
\usepackage{amsmath}
\usepackage{amsfonts}
\usepackage[colorlinks=true]{hyperref}
\usepackage{xcolor}
\usepackage{soul}
\usepackage{notes2bib}
\hypersetup{
    citecolor = blue,
	 linkcolor = blue
}

\author{Kuan-Sen Lin}
\affiliation[NTU Phys]
{Department of Physics, National Taiwan University, Taipei 10617, Taiwan}
\alsoaffiliation[UIUC]
{Department of Physics, University of Illinois at Urbana-Champaign, Urbana, IL, 61801-3080, USA}
\alsoaffiliation[IAMS]
{Institute of Atomic and Molecular Sciences, Academia Sinica, Taipei 10617, Taiwan}
\author{Mei-Yin Chou}
\email{mychou6@gate.sinica.edu.tw}
\phone{+886-2-2789-9900}
\affiliation[IAMS]
{Institute of Atomic and Molecular Sciences, Academia Sinica, Taipei 10617, Taiwan}
\alsoaffiliation[NTU Phys]
{Department of Physics, National Taiwan University, Taipei 10617, Taiwan}
\alsoaffiliation[Gatech]
{School of Physics, Georgia Institute of Technology, Atlanta, Georgia 30332, USA}

\title[An \textsf{achemso} demo]
  {Topological Properties of Gapped Graphene Nanoribbons with Spatial Symmetries}

\abbreviations{IR,NMR,UV}
\keywords{Graphene nanoribbons, geometrical phase, $\mathbb{Z}_{2}$ topological invariant, heterojunctions, localized state}

\begin{document}

\begin{tocentry}
\begin{center}
    \includegraphics[scale=1]{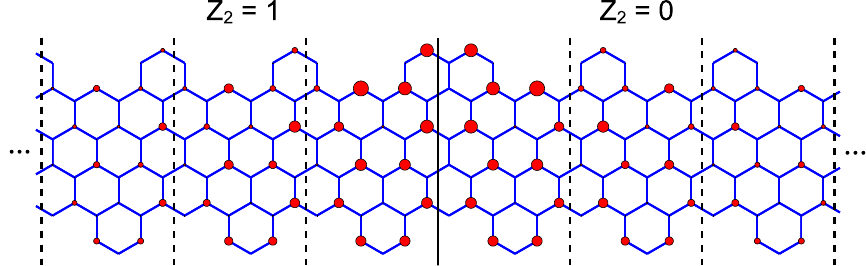}
\end{center}





\end{tocentry}

\clearpage
\begin{abstract}
  To date, almost all of the discussions on topological insulators (TIs) have focused on two- and three-dimensional systems. One-dimensional (1D) TIs manifested in real materials, in which localized spin states may exist at the end or near the junctions, have largely been unexplored. Previous studies have considered the system of gapped graphene nanoribbons (GNRs) possessing spatial symmetries (e.g. inversion) with only termination patterns commensurate with inversion- or mirror-symmetric unit cells. In this work, we prove that a symmetry-protected $\mathbb{Z}_{2}$ topological classification exists for any type of termination. In these cases the Berry phase summed up over all occupied bands turns out to be $\pi$-quantized in the presence of the chiral symmetry. However, it does not always provide the correct corresponding $\mathbb{Z}_{2}$  as one would have expected. We show that only the origin-independent part of the Berry phase gives the correct bulk-boundary correspondence by its $\pi$-quantized values. The resulting $\mathbb{Z}_{2}$ invariant depends on the choice of the 1D unit cell (defined by the nanoribbon termination) and is shown to be connected to the symmetry eigenvalues of the wave functions at the center and boundary of the Brillouin zone. Using the cove-edged GNRs as examples, we demonstrate the existence of localized states at the end of some GNR segments and at the junction between two GNRs based on a topological analysis. The current results are expected to shed light on the design of electronic devices based on GNRs as well as the understanding of the topological features in 1D systems.
\end{abstract}

\section{Keywords}
Graphene nanoribbons, geometrical phase, $\mathbb{Z}_{2}$ topological invariant, heterojunctions, localized state

\clearpage

 The topology of energy levels has been widely used in discussing novel phenomena in condensed matter physics, for example, the quantum Hall effect~\cite{QHE_exp,QHE_TKNN}, the quantum spin Hall effect~\cite{QSHE_Z2,QSHE_graphene}, surface/edge states~\cite{Zak_app_GNR,Zak_app_SS,Zak_app_Ca3P2,Zak_app_semimetal} and other quantum phases of matter~\cite{Topological_band_hasan,Zhang_Rev_Mod_Phys,Topological_band_Hsin_Lin,Mele_Rev_Mod_Phys,TCI_Fu,Chiu_reflection,Chiu_Rev_Mod_Phys,TCI_order2}.  Spatial symmetries are important factors in these classifications~\cite{Lee_arXiv,TCI_Fu,TCI_order2,Chiu_Rev_Mod_Phys,Chiu_reflection}; for example, the topological crystalline insulator is connected with the preserved rotational symmetry at the surface~\cite{TCI_Fu}. Most of the previous studies have focused on materials in two and three dimensions. Recently, the $\mathbb{Z}_{2}$ topological phases in semiconducting armchair graphene nanoribbons (aGNRs) have been investigated~\cite{CaoPRL,Louie_topological_band_engineering,engineering_robust_GNR}, which demonstrated that different widths and end shapes can lead to distinct topological properties and that new end and junction states can be successfully engineered. However, the discussions so far have been limited to GNRs with inversion- or mirror-symmetric unit cells. Given the fact that several complex one-dimensional (1D) GNRs have been synthesized using the bottom-up approach with atomically precise control in a variety of configurations~\cite{exp2,exp10,exp11,exp14,exp7,exp8,exp12,exp6,exp1,exp4}, a comprehensive study of their topological properties is of great interest.

In this work, we establish the theoretical description of these  gapped GNRs with spatial symmetries as 1D topological insulators for any type of termination by considering both spatial and chiral symmetries. We start from the Zak phase~\cite{zak} defined for a 1D system by
\begin{equation}
{\gamma _n} = i\int\limits_{ - \pi / d }^{\pi / d} {dk\left\langle {{u_{n,k}}} \right|{\partial _k}\left| {{u_{n,k}}} \right\rangle },
\label{zakintegral} 
\end{equation} 
where $u_{n,k} $ denotes the cell-periodic part of the Bloch wave function $\psi_{n,k} $ for band $n$ and crystal momentum $k$ with the length of the 1D lattice vector set as $d$, and the real-space integral is over one unit cell. The Zak phase is a geometrical phase acquired by a Bloch electron after an adiabatic transport across the whole Brillouin zone (BZ) and actually an open-path Berry phase in 1D~\cite{Resta_berry_phase}. Although the quantized Zak phase has been widely used to investigate the boundary states in 1D systems~\cite{CaoPRL} or higher dimensional systems with effective 1D Hamiltonians~\cite{Zak_app_GNR,Zak_app_Ca3P2,Zak_app_SS,Zak_app_semimetal}, the Zak phase is actually a quantity that depends on the choice of the real-space origin~\cite{Zak_origin_dependence,Intercellular}. Furthermore, since the space-integral in Eq.~\ref{zakintegral} involves the cell-periodic function $u_{n,k}$, $\gamma_{n}$ does not depend on the choice of the unit cell and therefore may not provide the connection with ribbon termination patterns. Thus, our first goal is to establish a {\it general} formulation to extract a unit-cell-dependent but origin-independent geometrical phase at half filling (called $\gamma_{2}$ below). By half filling we mean a neutral GNR with all valence states occupied. After getting the analytical expression of $\gamma_{2}$, we prove that it is $\pi$-quantized for an {\it inversion-symmetric unit cell}. By employing the tight-binding (TB) wave functions for gapped GNRs with the inversion symmetry, we further show that (1) in the presence of the chiral symmetry, $\gamma_{2}$ is always quantized to $\pi$ for an {\it arbitrary unit cell} at half filling, giving rise to a well-defined $\mathbb{Z}_{2}$ invariant and that (2) this $\mathbb{Z}_{2}$ invariant depends on the choice of the unit cell (namely, ribbon termination) and is connected to the parity of the wave functions at the center and the boundary of the BZ. In contrast, although the total Zak phase is also quantized to $\pi$ in this case with a proper choice of the origin, it is not a proper quantity to invoke the bulk-boundary correspondence.

After the theoretical formulation, we will use the cove-edged~\cite{exp10,exp11,exp12} GNRs as examples to demonstrate the aforementioned properties. We further show that even though higher-order hopping parameters break the chiral symmetry in graphene, these are small perturbations that do not eliminate the existence of localized states in gapped GNRs. In addition, possible extension of the present approach to gapped GNRs with other spatial symmetries will be discussed.

We focus on complex 1D systems which are periodic in only one direction (for example, $x$) with many atoms having finite coordinates in the other perpendicular directions (for example, $y$ and $z$). With the various configurations of synthesized GNRs~\cite{exp2,exp10,exp11,exp14,exp7,exp8,exp12,exp6,exp1,exp4}, the periodic unit cell is not necessarily rectangular. The basis of atoms in one unit cell is represented with respect to a real-space origin, which can be chosen independently of the unit cell choice. We assume that the spin-orbit coupling is negligible~\cite{SO} and focus on a spinless model for GNRs. By plugging $u_{n,k}\left( {\bf r} \right) = e^{-ikx} \psi_{n,k} \left( {\bf r} \right)$ into Eq.~\ref{zakintegral} we get $\gamma_{n} = \gamma_{1,n} + \gamma_{2,n}$, where
\begin{equation}
 \gamma_{1,n} = \int\limits_{ - \pi /d}^{\pi /d} {dk\left\langle {\psi_{n,k}} \right|x\left| {\psi_{n,k}} \right\rangle } , \label{gamma1}
\end{equation}
and
\begin{equation}
 \gamma_{2,n} = i\int\limits_{ - \pi /d}^{\pi /d} {dk\left\langle {\psi_{n,k}} \right|{\partial _k}\left| {\psi_{n,k}} \right\rangle }. \label{gamma2}
\end{equation}

Since $\gamma_{1,n}$ involves the expectation value of the $x$ coordinate for the Bloch wave function ${\psi _{n,k}}\left( {\bf{r}} \right)$, its value depends on the choice of the real-space origin~\cite{Zak_origin_dependence} and the unit cell. Actually, $\gamma_{1,n}$ is related to the classical electronic polarization, as shown in Ref.~\citenum{Intercellular}. On the other hand, the value of $\gamma_{2,n}$, as a pure $k$-space quantity based on the Bloch wave function, does not depend on the choice of the real-space origin. When we change the gauge of the global phase by $\psi_{n,k} \left( {\bf r} \right) \to \psi'_{n,k} \left( {\bf r} \right) = e^{i \theta_{n} \left( {k} \right) } \psi_{n,k} \left( {\bf r} \right)$, the value of $\gamma_{2,n}$ will change by $2p\pi$ where $p\in \mathbb{Z}$, as long as we require $\psi ' _{n,k} \left( {\bf r} \right) = \psi ' _{n,k+G} \left( {\bf r} \right)$ with $G=\frac{2\pi}{d}$. Therefore "$\gamma_{2,n}$ mod $2\pi$" is the essential part of $\gamma_{2,n}$. We note that $\gamma_{2,n}$ is equivalent to the intercellular Zak phase in Ref.~\citenum{Intercellular} up to a $2\pi$ ambiguity.

The expression of $\gamma_{2,n}$ can be greatly simplified if the system has certain symmetries. For example, if the Hamiltonian $H\left( {{\bf{r}},{\bf{p}}} \right) = \frac{{{{\bf{p}}^2}}}{{2m}} + V\left( {\bf{r}} \right)$ has the inversion symmetry, by taking the real-space origin at the GNR inversion center and assuming no degeneracy at general $k$-points \bibnote{It is possible to have degeneracies at the center and boundary of the BZ as dictated by the inversion symmetry. These will not affect our derivation. If there exists accidental degeneracies at general $k$-points, one can also possibly identify individual bands by a continuous gauge on the tight-binding wave functions going through the accidental crossing.}, we have ${e^{i{\phi _n}\left( k \right)}}{\psi _{n,k}}\left( { - {\bf{r}}} \right) = {\psi _{n, - k}}\left( {\bf{r}} \right)$, where ${{\phi _n}\left( k \right)}$ is a continuous phase dependent on $k$. At $\overline k = 0,\frac{\pi}{d}$ (the center and boundary of the BZ), using the property of the Bloch wave function that ${\psi _{n,k}}\left( {\bf{r}} \right) = {\psi _{n,k + G}}\left( {\bf{r}} \right)$ , we have
\begin{equation}
\widehat I{\psi _{n,\overline k }}\left( {\bf{r}} \right) = {e^{ - i{\phi _n}\left( {{\overline k}} \right)}}{\psi _{n,\overline k }}\left( {\bf{r}} \right) = {\zeta _n}\left( {\overline k } \right){\psi _{n,\overline k }}\left( {\bf{r}} \right).
\end{equation}
Here ${\psi _{n,\overline k }}\left( {\bf{r}} \right)$ is an eigenfunction of the inversion operator $\widehat I$ with an eigenvalue ${\zeta _n}\left( {{\overline k}} \right)$ of $+1$ or $-1$. 

Without loss of generality, let us consider a compact, namely without any separation, unit cell $\Omega = \Omega_{1} + \Omega_{2}$ and choose the GNR inversion center residing in $\Omega$ as the origin $O$. $\Omega_{1}$ is the maximal volume (within the unit cell $\Omega$) that maps to itself under the inversion upon $O$, and the rest volume is $\Omega_{2}$ (see Fig. S1 in Supporting Information). Summing up all the $\gamma_{2,n}$ of occupied bands to half filling, we obtain the origin-independent $\gamma_{2}$ as 
\begin{align}
{\gamma _{\rm{2}}} {\rm{ = }}\sum\limits_{n \in occ} {{\gamma _{2,n}}} & = \sum\limits_{n \in occ} \left( {{\phi _n}\left( {\frac{\pi }{d}} \right) - {\phi _n}\left( 0 \right) - d\int\limits_0^{\pi /d} {dk{Q_{2,n,k}}} } \right)  \\
& = \sum\limits_{n \in occ} {{\mathop{\rm Arg}\nolimits} \left( {{\zeta _n}\left( 0 \right){\zeta _n}\left( {\frac{\pi }{d}} \right)} \right) - \pi Q_{2} }, \label{derivation_1}
\end{align}
where ${Q_{2,n,k}} = \int\limits_{{\Omega _2}} {d{\bf{r}}{{\left| {\psi_{n,k} \left( {\bf r} \right)} \right|}^2}}$ measures the charge in $\Omega_{2}$ within the unit cell $\Omega$ from the Bloch wave function $\psi_{n,k} \left( {\bf r} \right)$, and $Q_{2}$ measures the total charge in $\Omega_{2}$ summed over the occupied bands (see Supporting Information for the details of the derivation). From the definitions of $\Omega_{1}$ and $\Omega_{2}$, we know that if the unit cell is itself inversion-symmetric, $\Omega_{2}$ will be zero, leading to $Q_{2} = 0$. For such a case we have 
\begin{equation}
\gamma_{2} = \sum\limits_{n \in occ} {{\rm Arg}\left( {{\zeta _n}\left( 0 \right){\zeta _n}\left( {\frac{\pi }{d}} \right)} \right)}.
\label{contiZ2}
\end{equation}
Since ${\zeta _n}\left( {\overline k } \right) \in \{ +1,-1 \}$, $\gamma_{2}$ must be quantized to $\pi$. We thus obtain a $\mathbb{Z}_{2}$ classification for an inversion-symmetric unit cell as
\begin{equation}
{\left( { - 1} \right)^{{\mathbb{Z}_2}}} = {e^{i{\gamma _2}}} = \prod\limits_{n \in occ} {{\zeta _n}\left( 0 \right){\zeta _n}\left( {\frac{\pi }{d}} \right)}.
\label{onlyinv}
\end{equation}
In this case, the $\mathbb{Z}_{2}$ invariant only depends on the parity of $\psi_{n,k} \left( {\bf r} \right)$ at the center and boundary of the BZ. The result is similar to those in two and three dimensions where in the presence of inversion symmetry the $\mathbb{Z}_{2}$ invariants can be determined from the knowledge of the parity of occupied states at the BZ center and boundaries~\cite{Zak_app_semimetal,Inversion_TI_Fu}. For a general unit cell with $\Omega_{2} \ne 0$, $\gamma_{2}$ is not necessarily quantized to $\pi$~\cite{Zak_app_SS,Intercellular,generalized_Zak}. As will be shown below, for gapped GNRs with the inversion symmetry, the chiral symmetry imposes an additional constraint leading to a $\pi$-quantized $\gamma_{2}$ for an arbitrary unit cell. 

We now consider the effect of chiral symmetry. We will use the tight-binding method for this part of the discussion. The Bloch wave function is represented as ${\psi _{n,k}}\left( {\bf{r}} \right) = \sum\limits_{l,m} {{\alpha _{n,l}}\left( k \right){e^{ikmd}}{\chi _{m,l}}\left( {\bf{r}} \right)} $, where ${{\chi _{m,l}}\left( {\bf{r}} \right)}$ is the atomic orbital $l$ in unit cell $m$. With this phase convention, the Bloch Hamiltonian $\mathcal{H}{\left( k \right)}$ and the coefficients ${{\alpha _{n,l}}\left( k \right)}$ are periodic in $k$-space with a periodicity of $G$. It can be shown~\cite{Intercellular} that, after mod $2\pi$, $\gamma_{2}$ can be written as
\begin{equation}
{\gamma _{2,n}} = i \sum\limits_{n \in occ} {\int\limits_{ - \pi /d}^{\pi /d} {dk{\left[ {{\psi _{n,k}}} \right] ^{\dag}}{\partial _k}{\left[ {{\psi _{n,k}}} \right]} }} ,
\label{rewritten}
\end{equation}
where $\left[ {{\psi _{n,k}}} \right] = {\left[ {\begin{array}{*{20}{c}}
{{\alpha _{n,1}}\left( k \right)}&{{\alpha _{n,2}}\left( k \right)}& \cdots &{{\alpha _{n,D - 1}}\left( k \right)}&{{\alpha _{n,D}}\left( k \right)}
\end{array}} \right]^T}$ is the eigenvector for band $n$ of the Bloch Hamiltonian $\mathcal{H} \left( {k} \right)$, and $D$ is the number of orthogonal atomic orbitals within one unit cell.

If we only consider the nearest-neighbor hopping in the tight-binding model, the Hamiltonian for GNRs will have the chiral symmetry. In this case, the components of $\left[ {{\psi _{n,k}}} \right]$ will satisfy $\sum\limits_{n \in occ} {{{\left| {{\alpha _{n,l}}\left( k \right)} \right|}^2}} = 0.5$ $\forall $ $l,k$ (see Supporting Information). If we denote $\gamma_{2}$ and $\gamma'_{2}$ as the results for two different unit cells in the {\it same} periodic 1D ribbon, $\gamma_{2}$ and $\gamma'_{2}$ will satisfy $\gamma'_{2} = \gamma_{2} - M\pi $ (see Supporting Information), where $M$ $\in$ $\mathbb{Z}$ and depends on the shuffling of the carbon sites when the definition of the unit cell changes. It is noted that imposing the chiral symmetry will also determine the change of the origin-dependent part of Zak phase summed up to half filling as $\gamma_{1}' = \gamma_{1} + M\pi$ (see Supporting Information). Therefore $\gamma = \gamma_{1} + \gamma_{2}$ will be invariant upon the choice of the unit cell in the same periodic 1D ribbon owing to a $M\pi$ phase shift between $\gamma_{1}$ and $\gamma_{2}$. 

For a gapped GNR with both inversion and chiral symmetries, the numbers of sites belonging to the A and B sublattices are the same~\cite{chiral1}. We can always start with an unit cell that exhibits a complete inversion symmetry ($\Omega_{2}=0$) which makes $\gamma_{2}$ quantized to $\pi$. If we further choose the real-space origin at this inversion center, we have $\gamma_{1}  = 0$ according to Eq.~\ref{gamma1}. It then follows that $\gamma_{1}$, $\gamma_{2}$ and $\gamma$ are all quantized to $\pi$. Using $\left( {-1} \right) ^{\mathbb{Z}_{2}} = e^{i \gamma_{2}}$, a different choice of the unit cell in the same GNR will give $\mathbb{Z}'_{2} = \left( {\mathbb{Z}_{2} - M} \right)$ mod $2$. Therefore, if $M$ is an odd (even) integer, the two unit cells of the same GNR are topologically distinct (equivalent). The generalized $\mathbb{Z}_{2}$ classification here is symmetry-protected.

When a long GNR segment is truncated, the atomic configuration at the end uniquely defines a unit cell in the remaining semi-infinite segment. According to the bulk-boundary correspondence, if the corresponding $\mathbb{Z}_{2}$ equals to $1$, there will be one or an odd number of localized end states at the termination with a mid-gap energy. That is, at least one localized state is protected. Also, if a heterojunction is formed from two long segments of GNRs whose corresponding 1D unit cells are topologically distinct (equivalent), there will be one or an odd (zero or an even) number of localized junction states at the boundary.

For the symmetry class of BDI with time-reversal symmetry, particle-hole symmetry, and sublattice (chiral) symmetry, a topologically non-trivial ground state exists only in one dimension~\cite{Schnyder_2008}, and the space of these quantum ground states is partitioned into topological sectors labeled by an integer ($\mathbb{Z}$). GNRs generally fall into this symmetry class. If the GNRs also possess additional spatial symmetries, as we have discussed in this Letter, a proper $\mathbb{Z}_{2}$ classification can be found by using the appropriate topological invariant. However we note that the winding number~\cite{winding_wen} can also classify the half-filling topological phases and can be calculated easily in the tight-binding scheme~\cite{Koshino_chiral}. We have checked the results for all examples in this work. The predictions from the winding number and the $\mathbb{Z}_{2}$ topological invariant deduced from the origin-independent Zak phase ($\gamma_{2}$) are consistent. The intertwined linkage between the original BDI $\mathbb{Z}$ classification and newly emergent $\mathbb{Z}_{2}$ classification in the presence of spatial symmetries in GNRs may be an interesting topic for future study.

To realize stable insulating phases suitable for electronic device applications, the band gap should be at least one order of magnitude larger than the thermal energy at room temperature ($\sim 0.025$ eV). For the cove-edged GNRs in the following topological analysis, the band gap from tight-binding calculations is approximately  $0.68$/$0.55$/$0.39$ eV for the width of $0.92$/$1.14$/$1.56$ nm (see Supporting Information).

\begin{figure}
  \includegraphics[scale=0.45]{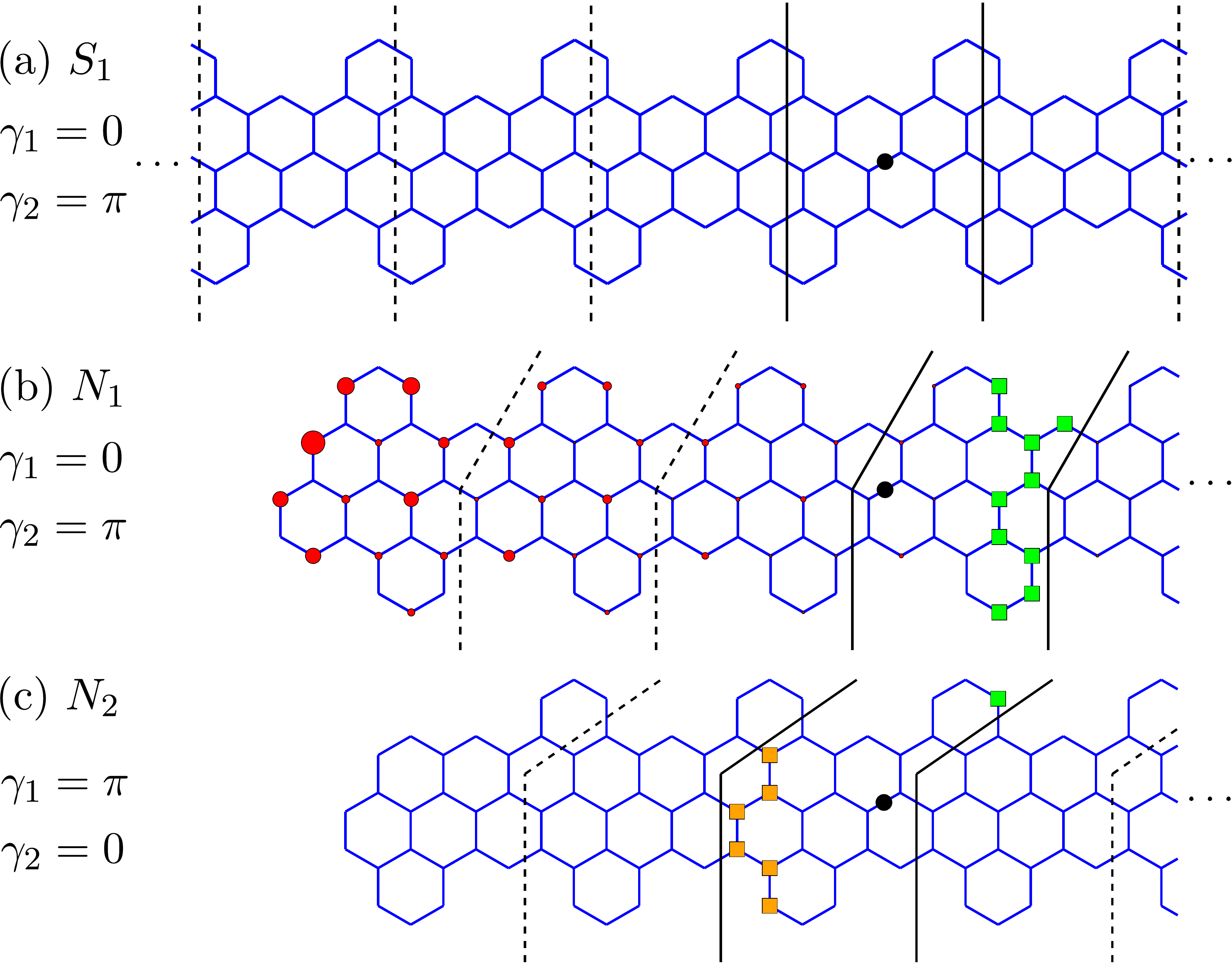}
  \caption{Different unit cell choices in the same GNR that has the inversion symmetry as indicated by the black lines in (a) - (c) with their corresponding Zak phases $\gamma_{1}$ and $\gamma_{2}$ shown (see text). The black circle is the real-space origin which is also an inversion center used to derive the $\gamma_{1}$ for unit cells in (a) - (c). In (a) the unit cell is inversion-symmetric while the unit cells in (b) and (c) are not, but their $\mathbb{Z}_{2}$ invariants can be obtained by counting the numbers of the green and orange squares in (b) and (c) (see text). The green and orange squares indicate the different atoms included in (b) and (c) compared with (a). The red circles in (b) show the relative probability distributions of the end state with $E=0$ and their areas proportional to the probabilities. The total area of the red circles is chosen to be $\pi \left( {0.75a_{0}} \right)^{2}$, where $a_{0}$ is the nearest-neighbor distance between carbon atoms.}
  \label{new_figure_1}
\end{figure}

We show in Fig.~\ref{new_figure_1} three different unit cells for the same cove-edged GNR~\cite{exp10,exp11,exp12} which satisfies the requirement of having a band gap and the inversion symmetry. In Fig.~\ref{new_figure_1} (a) the unit cell is inversion-symmetric and has $\mathbb{Z}_{2} = 1$ according to Eq.~\ref{onlyinv}. When changing from unit cell (a) to unit cell (b), both indicated by the black solid lines, ten carbon sites will be removed and included again marked by green squares in the next unit cell ($+1$ lattice vector). Therefore we have $M=+10$ in this case, but the essential difference of $\gamma_{2}$ between the two unit cells in Figs.~\ref{new_figure_1} (a) and (b) will be $0$. Similarly, when changing from unit cell (a) to (c), one carbon sites will be included from the next unit cell, while six carbon sites marked by orange squares will be taken from the left unit cell ($-1$ lattice vector), leading to $M=-5$ in this case. Hence, the essential difference of $\gamma_{2}$ between the two unit cells in Figs.~\ref{new_figure_1} (a) and (c) will be $\pi$. 

\begin{figure}
  \includegraphics[width=\textwidth]{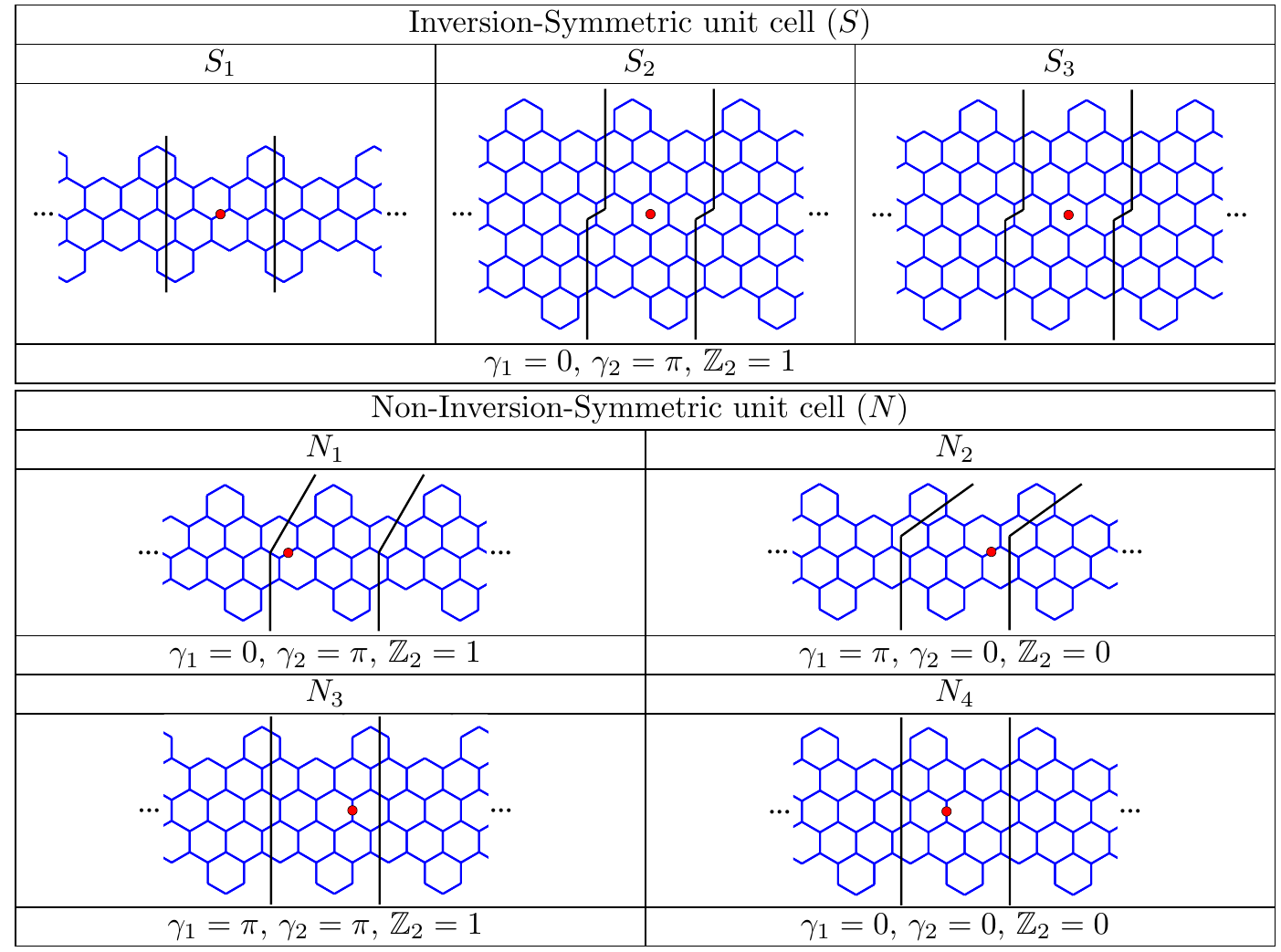}
  \caption{Different unit cells of cove-edged GNRs and their corresponding $\gamma_{1}$, $\gamma_{2}$ and $\mathbb{Z}_{2}$ invariants (see text). The red circles indicate the real-space origins used to derive $\gamma_{1}$, and they are also inversion centers for the GNRs. All the phases are presented after mod $2\pi$.}
  \label{Table1}
\end{figure}

We have performed tight-binding calculations with a nearest-neighbor interaction of $t_{1} \approx -2.7$ eV~\cite{grapheneTB} using the exact diagonalization method with 100 unit cells for cove-edged GNRs with two different truncation patterns in Figs.~\ref{new_figure_1} (b) and (c). The probability distribution of localized end states (if any) at the left termination is plotted on the carbon sites. The results confirm that there is {\it one} localized end state in Fig.~\ref{new_figure_1} (b) while there is {\it no} localized end state in Fig.~\ref{new_figure_1} (c). In a neutral GNR, the end state in Fig.~\ref{new_figure_1} (b) is singly occupied, leading to a localized spin state. The calculated unit-cell-dependent $\mathbb{Z}_{2}$ invariants for a few cove-edged GNR examples with a band gap and the inversion symmetry are shown in Fig.~\ref{Table1}. A choice of the origin at one of the inversion centers will simplify the calculation of $\gamma_{1}$. We note that both $\gamma_{1}$ and $\gamma_{2}$ are quantized to $\pi$ (and so is $\gamma=\gamma_{1}+\gamma_{2}$), but the value of $\gamma_{2}$ is used to define the topological quantity $\mathbb{Z}_{2}$.

\begin{figure}
  \includegraphics[scale=0.45]{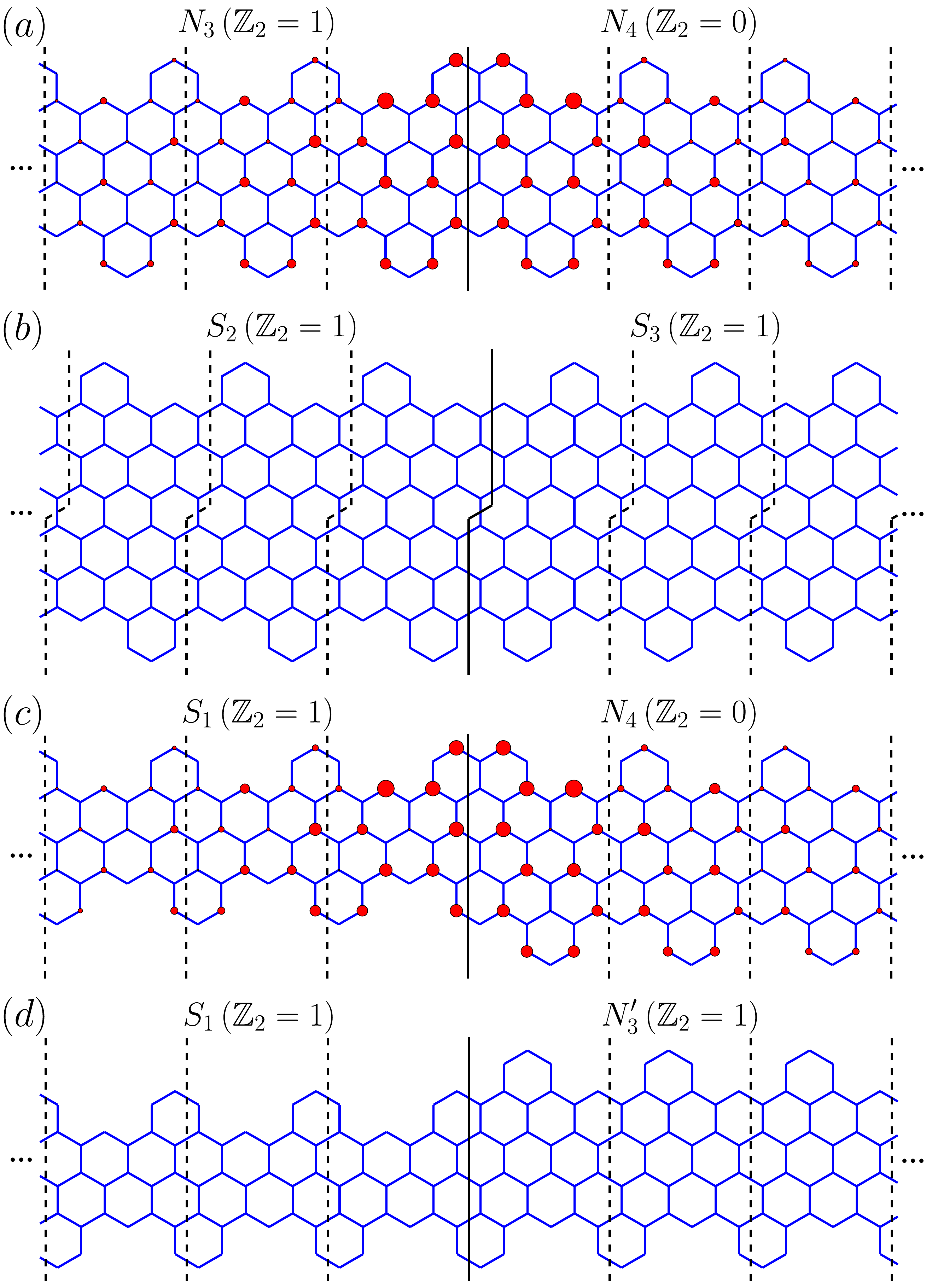}
  \caption{Heterojunctions formed by combining topologically distinct [(a) and (c)] and topologically equivalent [(b) and (d)] cove-edged GNR segments. The black solid lines denotes the interfaces between the GNR segments. The dashed lines denote the unit cells in each GNR segment. The $\mathbb{Z}_{2}$ invariant for each unit cell is taken from Fig.~\ref{Table1} and also shown in (a) - (d). The $N_{3}'$ GNR in (d) is equivalent to $N_{3}$ in Fig.~\ref{Table1} but with an additional $180^{\circ}$ rotation. The red circles in (a) and (c) show the relative probability distributions of the junction states with $E=0$, and their total area is chosen to be $\pi \left( {1.5a_{0}} \right)^{2}$, where $a_{0}$ is the nearest-neighbor distance between carbon atoms.}
  \label{new_picture_2}
\end{figure}

In Figs.~\ref{new_picture_2} (a) and (b), we show two heterojunctions formed from cove-edged GNR segments with the same width. Black solid lines indicate the domain boundaries, while dashed lines mark the unit cells for the left and right cove-edged GNR segments. According to Fig.~\ref{Table1}, the left segment in Fig.~\ref{new_picture_2} (a) has $\mathbb{Z}_{2} = 1$ (type $N_3$) while the right one has $\mathbb{Z}_{2} = 0$ (type $N_4$). In Fig.~\ref{new_picture_2} (b), the left and right segments both have $\mathbb{Z}_{2} = 1$ (types $S_{2}$ and $S_3$). We again perform tight-binding calculations to verify the existence of the junction states, using the same nearest-neighbor Hamiltonian as in the previous end-state calculations and 100 unit cells for both the left and right segments. As expected, our tight-binding results show that there is {\it one} localized junction state in Fig.~\ref{new_picture_2} (a) for which the probability distribution is plotted, while there is {\it no} localized junction state in Fig.~\ref{new_picture_2} (b).

We have constructed more heterojunctions from cove-edged GNRs with different widths in Figs.~\ref{new_picture_2} (c) and (d), and performed tight-binding calculations. The results show {\it one} localized junction state in Fig.~\ref{new_picture_2} (c) for which the probability distribution is plotted whereas there is {\it no} localized junction states in Fig.~\ref{new_picture_2} (d), consistent with the predictions based on their $\mathbb{Z}_{2}$ values deduced from $\gamma_{2}$.


Higher-order hoppings will break the exact chiral symmetry in graphene~\cite{chiral1}. When we include the second-nearest-neighbor hopping~\cite{grapheneTB} in our tight-binding model, $\gamma_{2}$ will not be exactly quantized to $\pi$ any more for cove-edged GNRs, but the band structure is only slightly modified, since the higher-order hoppings are significantly smaller than the nearest-neighbor one and can be regarded as perturbations. Thus the topological properties of real gapped GNRs may be deduced from the chiral-symmetric models. To further check this, we use a Slater-Koster-type formula \cite{Slater,TBG_1,TBG_2}, $ t\left( {r} \right) = -V_{pp\pi }^0\exp ( - \frac{{r - {a_{0}}}}{{{r_{0}}}})$, to parametrize all the possible hoppings at distance $r$, where $V_{pp\pi }^0 \approx 2.7$ eV, $a_{0} \approx 0.142$ nm is the nearest-neighbor distance between carbon atoms, and $r_{0} \approx 0.184\ a_{0}$ is a fitting parameter based on the energy bands obtained from first-principles calculations \cite{TBG_1,TBG_2}. We have performed full tight-binding calculations for examples present in this paper. The absence and presence of localized end and junction states are preserved, and their charge distributions are nearly the same as those obtained with only the nearest-neighbor hopping. This demonstrates the robustness of the topological properties derived from the chiral-symmetric Hamiltonian.


For gapped GNRs with other spatial symmetries, the $\pi$-quantized $\gamma_{2}$ can be formulated by the general procedure as follows. We first identify the spatial symmetry we would like to consider and denote its matrix representation acting on the Bloch states as $\mathcal{O} \left( k \right)$. The $\mathcal{O}$-symmetry is assumed to be order-$2$, thus $\mathcal{O} \left( {k} \right)$ can be inversion, reflection or even a $180^{\circ}$ rotational symmetry. We also assume that $\mathcal{O} \left( {k} \right)$ will flip the crystal momentum $k$ of the gapped Bloch Hamiltonian $\mathcal{H} \left( {k} \right)$ by $\mathcal{O} \left( {k} \right) \mathcal{H} \left( {k} \right) \mathcal{O} \left( {k} \right)^{-1} = \mathcal{H} \left( {-k} \right)$. We then find an $\mathcal{O}$-symmetric unit cell and calculate the $\mathcal{O}$-symmetry eigenvalues of the Bloch wave functions at $k=0,\frac{\pi}{d}$. By using 
\begin{equation}
{\left( { - 1} \right)^{{\mathbb{Z}_2}}} =e^{i \gamma_{2}} = {e^{i\sum\limits_{n \in occ} {{\rm Arg}\left( {{\xi _n}\left( 0 \right){\xi _n}\left( {\frac{\pi}{d}} \right)} \right)} }} = \prod\limits_{n \in occ} {{\xi _n}\left( 0 \right){\xi _n}\left( {\frac{\pi }{d}} \right)} ,
\end{equation}
where $\xi_{n} \left( {0} \right)$ and $\xi_{n} \left( {\frac{\pi}{d}} \right)$ $\in \{ -1,+1\}$ denote the $\mathcal{O}$-symmetry eigenvalues of the Bloch wave functions at $k=0$ and $\frac{\pi}{d}$ respectively, we can get the $\mathbb{Z}_{2}$ invariant for this $\mathcal{O}$-symmetric unit cell. Then, using $\mathbb{Z}'_{2} = \left( {\mathbb{Z}_{2} - M} \right)$ mod $2$, we can obtain the $\mathbb{Z}_{2}$ invariant of the other non-$\mathcal{O}$-symmetric unit cell of the same GNR. Therefore, our discussions on $\pi$-quantized $\gamma_{2}$ can be applied to any relevant spatial symmetries in gapped GNRs. With such a general procedure, we expect that other types of cove-edged GNRs~\cite{exp10,exp11,exp12} or its equivalent structures~\cite{zz1212}, chiral GNRs~\cite{exp14,Gap_chiral_GNR}, armchair GNRs~\cite{exp2,exp7,exp8}, GNR multilayers~\cite{AB_bilayer1}, chevron GNRs~\cite{exp2,exp4,exp6}, and carbon nanotubes~\cite{CNT1} will all support well-defined $\mathbb{Z}_{2}$ invariants derived from the $\pi$-quantized $\gamma_{2}$ for arbitrary unit cells.

In conclusion, we present a complete formulation of the origin-independent Zak phase $\gamma_{2}$ for gapped GNRs with spatial symmetries, which is the relevant quantity describing the band topology in one dimension. Using the tight-binding method, we show that $\gamma_{2}$ is always quantized to $\pi$ for arbitrary unit cells of gapped GNRs with the inversion and chiral symmetries, leading to a well-defined $\mathbb{Z}_{2}$ invariant. Such unit-cell-dependent $\mathbb{Z}_{2}$ invariants can dictate the boundary phenomena according to the bulk-boundary correspondence, as demonstrated by using cove-edged GNRs as examples. Although the total Zak phase $\gamma$ could also be quantized to $\pi$ with a proper choice of the origin, it does not provide a consistent prediction on the emergence of localized end or junction states. We have also confirmed the robustness of the localized end or junction states by adding higher-order hoppings that break the chiral symmetry. In addition, the present treatment can be extended to handle other order-$2$ spatial symmetries such as the reflection and rotational symmetries. This will lead to potential applications in the design of complex 1D heterojunctions with localized spin-polarized junction states at the mid-gap.

\begin{acknowledgement}

The authors thank Drs. J. Zak, J. W. Rhim, Y.-H. Chan and C.-K. Chiu for helpful discussions. This work is supported by a Thematic Project (AS-TP-106-M07) at Academia Sinica.

\end{acknowledgement}

\begin{suppinfo}

The supporting information contains detailed derivations on i) $\gamma_{2} = \sum\limits_{n \in occ} {{\mathop{\rm Arg}\nolimits} \left( {{\zeta _n}\left( 0 \right){\zeta _n}\left( {\frac{\pi }{d}} \right)} \right) - \pi Q_{2} }$, ii) $\sum\limits_{n \in occ} {{{\left| {{\alpha _{n,l}}\left( k \right)} \right|}^2}} = 0.5$ $\forall $ $l,k$,  iii) $\gamma'_{2} = \gamma_{2} - M\pi $ and $\gamma'_{1} = \gamma_{1} + M\pi $, and iv) the band structures of the cove-edged GNRs considered in the main text.

\end{suppinfo}

\clearpage


\bibliography{achemso-demo}

\begin{mcitethebibliography}{51}
\bibitem[Not()]{Note-1}
It is possible to have degeneracies at the center and boundary of the BZ as
  dictated by the inversion symmetry. These will not affect our derivation. If
  there exists accidental degeneracies at general $k$-points, one can also
  possibly identify individual bands by a continuous gauge on the tight-binding
  wave functions going through the accidental crossing.\relax
\mciteBstWouldAddEndPunctfalse
\mciteSetBstMidEndSepPunct{\mcitedefaultmidpunct}
{}{\mcitedefaultseppunct}\relax
\EndOfBibitem
\end{mcitethebibliography}

\end{document}


\section{i) Derivation of $ {\gamma _{\rm{2}}} = \sum\limits_{n \in occ} {{\mathop{\rm Arg}\nolimits} \left( {{\zeta _n}\left( 0 \right){\zeta _n}\left( {\frac{\pi }{d}} \right)} \right) - \pi Q_{2} } $}

\begin{figure}
  \includegraphics[scale=0.4]{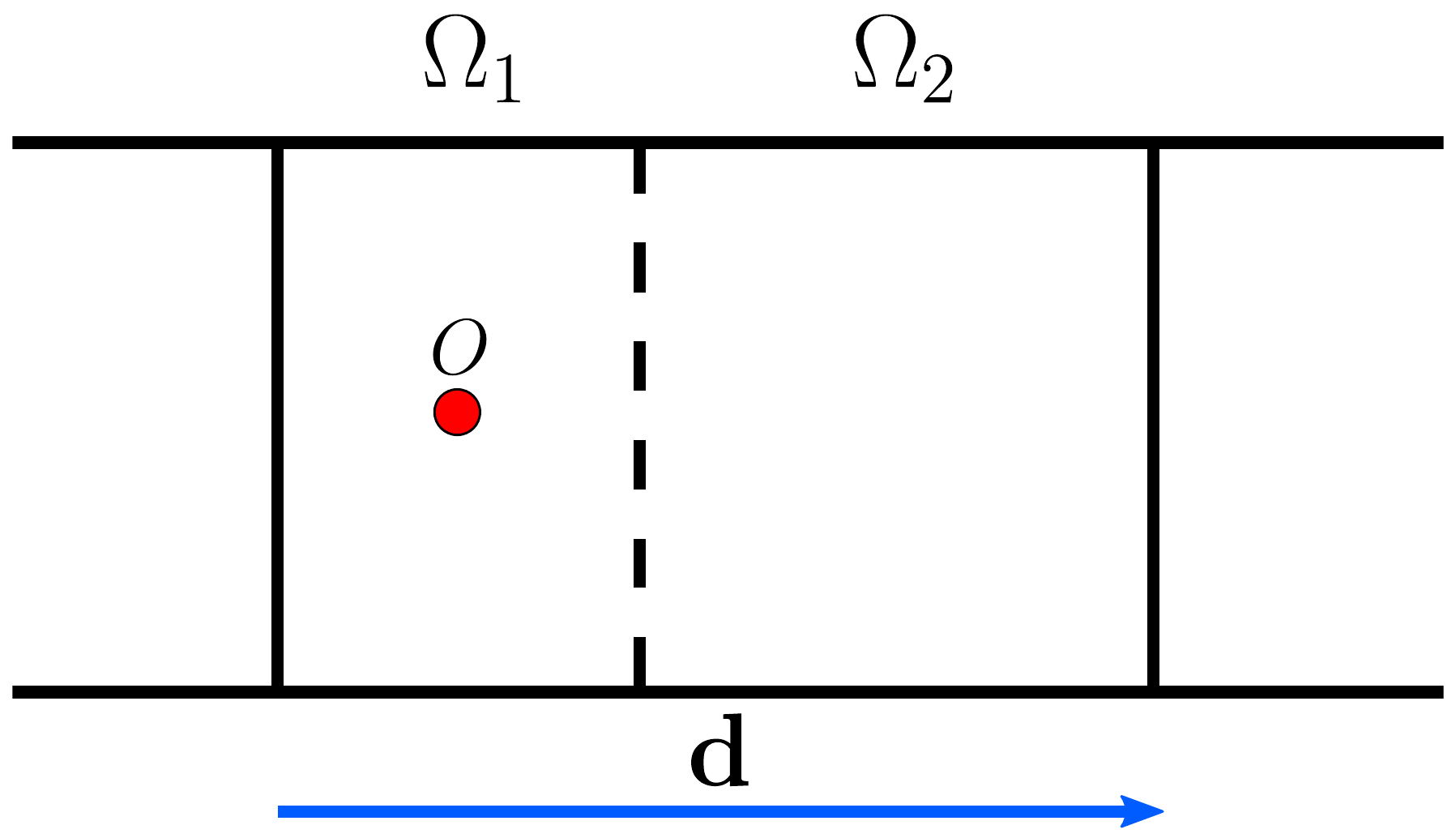}
  \caption{Unit cell $\Omega$ is a sum of two parts: $\Omega_{1}$ and $\Omega_{2}$. The red circle ($O$) is the inversion center, and $\Omega_{1}$ is invariant under the inversion.}
  \label{Derivation3}
\end{figure}

As explained in the main text, the origin-independent part of the Zak phase of band $n$ is given by
\begin{equation}
\gamma_{2,n} = i\int\limits_{ - \pi /d}^{\pi /d} {dk\int\limits_{{\Omega}} {d{\bf{r}}\psi _{n,k}^*\left( {\bf{r}} \right){\partial _k}\psi _{n,k}\left( {\bf{r}} \right)} }. \label{gamma_2_si}
\end{equation} 
Without loss of generality, we (1) choose the real-space origin $O$ at the GNR inversion center since the choice of the real-space origin will not affect the value of $\gamma_{2,n}$ and (2) let the origin reside in a compact unit cell $\Omega$ chosen to be $\Omega = \Omega_{1} + \Omega_{2}$, as shown in Fig.~\ref{Derivation3}. $\Omega_{1}$ is the maximal volume within the unit cell $\Omega$ that maps to itself under the inversion upon $O$, and the rest volume is $\Omega_{2}$. $\Omega_1$ and $\Omega_{2}$ are not necessarily rectangular for a particular GNR. We can write $\gamma_{2,n}$ in Eq.~\ref{gamma_2_si} as a sum of integrals over $\Omega_{1}$ and $\Omega_{2}$. For the former we have
\begin{align}
& i\int\limits_{ - \pi /d}^{\pi /d} {dk\int\limits_{{\Omega _1}} {d{\bf{r}}\psi _{n,k}^*\left( {\bf{r}} \right){\partial _k}\psi _{n,k}\left( {\bf{r}} \right)} }  \\
&= i\int\limits_0^{\pi /d} {dk\left( {\int\limits_{{\Omega _1}} {d{\bf{r}}\psi _{n,k}^*\left( {\bf{r}} \right){\partial _k}\psi _{n,k}\left( {\bf{r}} \right)}  - \int\limits_{{\Omega _1}} {d{\bf{r}}\psi _{n,-k}^*\left( {\bf{r}} \right){\partial _k}\psi _{n,-k}\left( {\bf{r}} \right)} } \right)} \label{eq1} \\
&  =  i\int\limits_0^{\pi /d} {dk\int\limits_{{\Omega _1}} {d{\bf{r}}\psi _{n,k}^*\left( {\bf{r}} \right){\partial _k}\psi _{n,k}\left( {\bf{r}} \right)}   }  - i\int\limits_0^{\pi /d} {dk\int\limits_{{\Omega _1}} {d{\bf{r}}{e^{ - i{\phi _n}\left( k \right)}}\psi _{n,k}^*\left( {- {\bf{r}}} \right) {{\partial _k} \left[ {e^{i{\phi _n}\left( k \right)}}\psi _{n,k}\left( {- {\bf{r}}} \right)\right]}  } }   \\
&  = \int\limits_0^{\pi /d} {dk{\partial _k}{\phi _n}\left( k \right){Q_{1,n,k}}}\ ,  \label{eq2}
\end{align}
where we have used ${e^{i{\phi _n}\left( k \right)}}{\psi _{n,k}}\left( { - {\bf{r}}} \right) = {\psi _{n, - k}}\left( {\bf{r}} \right)$ in the main text and the property that $\Omega_{1}$ maps to itself, and have set
\begin{equation}
{Q_{1,n,k}} = \int\limits_{{\Omega _1}} {d{\bf{r}}{{\left| {\psi _{n,k}\left( {\bf{r}} \right)} \right|}^2}}.
\end{equation}
As we can see, $Q_{1,n,k}$ is proportional to the charge in $\Omega_1$ contributed from ${\psi _{n,k}\left( {  {\bf{r}}} \right)}$ with ${\psi _{n,k}\left( {  {\bf{r}}} \right)}$ properly normalized within one unit cell.

To calculate the integral over $\Omega_{2}$, we note that for ${\bf r} \in \Omega_{2}$ in Fig.~\ref{Derivation3}, $- {\bf r} + {\bf d}$ will map back to $\Omega_{2}$, and we have
\begin{equation}
\psi _{n,k}\left( { - {\bf{r}} + {\bf{d}}} \right) = {e^{ikd}}\psi _{n,k}\left( { - {\bf{r}}} \right) = {e^{ikd}}{e^{ - i{\phi _n}\left( k \right)}}\psi _{n,-k}\left( {\bf{r}} \right).
\end{equation}
The contribution from $\Omega_{2}$ to $\gamma_{2,n}$ is
\begin{align}
& i\int\limits_{ - \pi /d}^{\pi /d} {dk\int\limits_{{\Omega _2}} {d{\bf{r}}\psi _{n,k}^*\left( {\bf{r}} \right){\partial _k}\psi _{n,k}\left( {\bf{r}} \right)} }  \\
& = i\int\limits_0^{\pi /d} {dk\left( {\int\limits_{{\Omega _2}} {d{\bf{r}}\psi _{n,k}^*\left( {\bf{r}} \right){\partial _k}\psi _{n,k}\left( {\bf{r}} \right)}  - \int\limits_{{\Omega _2}} {d{\bf{r}}\psi _{n,-k}^*\left( {\bf{r}} \right){\partial _k}\psi _{n,-k}\left( {\bf{r}} \right)} } \right)}  \\
& = i\int\limits_0^{\pi /d} {dk {\int\limits_{{\Omega _2}} {d{\bf{r}}\psi _{n,k}^*\left( {\bf{r}} \right){\partial _k}\psi _{n,k}\left( {\bf{r}} \right)}  } } \\
&\ \ \  - i\int\limits_0^{\pi /d} {dk\int\limits_{{\Omega _2}} {d{\bf{r}}{e^{ikd}}{e^{ - i{\phi _n}\left( k \right)}}\psi _{n,k}^*\left( {  - {\bf r} + {\bf d} } \right){{\partial _k} \left[ {e^{ - ikd}}{e^{i{\phi _n}\left( k \right)}}\psi _{n,k}\left( { - {\bf r} + {\bf d}} \right) \right]}  }   }   \\
& =  \int\limits_0^{\pi /d} {dk\int\limits_{{\Omega _2}} {d{\bf{r}}\left( { {{\left| {\psi _{n,k}\left( {{\bf{r}}} \right)} \right|}^2  {{\partial _k}{\phi _n}\left( k \right)} } - d{{\left| {\psi _{n,k}\left( {{\bf{r}}} \right)} \right|}^2}} \right)} } \\
& =    \int\limits_0^{\pi /d} {dk{Q_{2,n,k}} {\partial _k}{\phi _n}\left( k \right)}  - d\int\limits_0^{\pi /d} {dk{Q_{2,n,k}}}\ ,  \label{Omega2partialresult} 
\end{align}
where we have set
\begin{equation}
{Q_{2,n,k}} = \int\limits_{{\Omega _2}} {d{\bf{r}}{{\left| {\psi _{{n,k}}^{}\left( {\bf{r}} \right)} \right|}^2}} .
\end{equation}
We have again used the mapping property of $\Omega_{2}$ to change the argument of ${\psi _{n,k}}\left( {\bf{r}} \right)$ in the integrand  from $ - {\bf r} + {\bf d}$ to $\bf r$ without changing the real-space integral range for $\Omega_{2}$. 

Combining Eqs.~\ref{eq2} and \ref{Omega2partialresult}, we get
\begin{align}
\gamma_{2,n} & =  i\int\limits_{ - \pi /d}^{\pi /d} {dk\int\limits_\Omega  {d{\bf{r}}\psi _{_{n,k}}^*\left( {\bf{r}} \right){\partial _k}\psi _{_{n,k}}^{}\left( {\bf{r}} \right)} }  = \int\limits_0^{\pi /d} {dk\left( {{Q_{1,n,k}} + {Q_{2,n,k}}} \right) {\partial _k}{\phi _n}\left( k \right)}  - d\int\limits_0^{\pi /d} {dk{Q_{2,n,k}}} \label{addQbefore} \\
&  = \int\limits_0^{\pi /d} {dk{\partial _k}{\phi _n}\left( k \right)}  - d\int\limits_0^{\pi /d} {dk{Q_{2,n,k}}}  = \left[ {{\phi _n}\left( {\frac{\pi }{d}} \right) - {\phi _n}\left( 0 \right)}\right] - d\int\limits_0^{\pi /d} {dk{Q_{2,n,k}}}\ , \label{result}  
\end{align}
where we have used
\begin{equation}
 {Q_{{\rm{1}},n,k}}{\rm{ + }}{Q_{2,n,k}} = \int\limits_{{\Omega _{\rm{1}}}} {d{\bf{r}}{{\left| {\psi _{{n,k}}^{}\left( {\bf{r}} \right)} \right|}^2}} {\rm{ + }}\int\limits_{{\Omega _2}} {d{\bf{r}}{{\left| {\psi _{{n,k}}^{}\left( {\bf{r}} \right)} \right|}^2}}  = 1 .
\end{equation}
If we do a gauge transformation ${\psi _{n,k}}\left( {\bf{r}} \right) \to {\psi' _{n,k}}\left( {\bf{r}} \right) = {e^{i{\theta _n}\left( k \right)}}{\psi _{n,k}}\left( {\bf{r}} \right)$ where ${{\theta _n}\left( k \right)}$ is a continuous phase, then ${\gamma '_{2,n}} = {\gamma _{2,n}} + \left( {{\theta _n}\left( {\frac{\pi }{d}} \right) - {\theta _n}\left( { - \frac{\pi }{d}} \right)} \right)$. However, to satisfy ${\psi '_{n,k}}\left( {\bf{r}} \right) = {\psi '_{n,k + G}}\left( {\bf{r}} \right)$, the continuous phase should satisfy ${\theta _n}\left( {\frac{\pi }{d}} \right) - {\theta _n}\left( { - \frac{\pi }{d}} \right) = 2p\pi$, where $p \in \mathbb{Z}$. Therefore, under a gauge transformation on the global phase, $\gamma_{2,n}$ can have a $2\pi$ ambiguity, so the essential part of $\gamma_{2,n}$ is "$\gamma_{2,n}$ mod $2\pi$". Furthermore, at $ \overline k =0$ and $\frac{\pi}{d}$, we have $e^{-i\phi_{n}\left( {\overline k } \right)} = \zeta \left( {\overline k} \right)$, where $\zeta \left( {\overline k} \right) \in \{ {+1,-1} \}$ is the inversion eigenvalue of the Bloch wave function $\psi_{n,k} \left( {\bf r} \right)$ with respect to the inversion center. We thus write
\begin{equation}
{\phi _n}\left( {\frac{\pi }{d}} \right) - {\phi _n}\left( 0 \right) = {\rm Arg}\left( {{\zeta _n}\left( 0 \right){\zeta _n}\left( {\frac{\pi }{d}} \right)} \right).
\end{equation}

We can further simplify the last term of Eq.~\ref{result}. Using ${e^{i{\phi _n}\left( k \right)}}{\psi _{n,k}}\left( { - {\bf{r}}} \right) = {\psi _{n, - k}}\left( {\bf{r}} \right)$ and $\rho_{n,k} \left( {\bf r} \right) = | \psi_{n,k} \left( {\bf r} \right) |^{2}$, we have $\rho_{n,k} \left( {\bf - r} \right) = \rho_{n,-k} \left( {\bf r} \right)$, or equivalently $\rho_{n,k} \left( {\bf  r} \right) = \rho_{n,-k} \left( {\bf -r} \right)$. Therefore, 
\begin{align}
{Q_{2,n,k}}  & =  \int\limits_{{\Omega _2}} d{\bf r} \rho_{n,k} \left( {\bf r} \right) = \int\limits_{{\Omega _2}} d{\bf r} \rho_{n,-k} \left( {\bf -r} \right) \\
& = \int\limits_{{\Omega _2}} d{\bf r} \rho_{n,-k} \left( {\bf -r + \bf d} \right) \label{change1}  \\
& = \int\limits_{{\Omega _2}} d{\bf r}' \rho_{n,-k} \left( {\bf r }' \right) = {Q_{2,n,-k}} , 
\label{change2}
\end{align}
where we have used the property that under the transformation ${\bf r} \to {\bf r}' = {\bf -r} + \bf d$ the integral range actually does not change in passing from Eq.~\ref{change1} to \ref{change2}.

Summing up occupied bands to half filling for the insulating ground state we have
\begin{align}
\gamma_{2} & = \sum\limits_{n \in occ} {\gamma_{2,n}}   = \sum\limits_{n \in occ} {{\rm Arg}\left( {{\zeta _n}\left( 0 \right){\zeta _n}\left( {\frac{\pi }{d}} \right)} \right)}  - d\int\limits_0^{\pi /d} {dk\sum\limits_{n \in occ} {{Q_{2,n,k}}} } \\
& = \sum\limits_{n \in occ} {{\rm Arg}\left( {{\zeta _n}\left( 0 \right){\zeta _n}\left( {\frac{\pi }{d}} \right)} \right)} - \frac{d}{2}\int_0^{\pi /d} {dk\sum\limits_{n \in occ} {\left( {{Q_{2,n,k}} + {Q_{2,n, - k}}} \right)} }\\
& = \sum\limits_{n \in occ} {{\rm Arg}\left( {{\zeta _n}\left( 0 \right){\zeta _n}\left( {\frac{\pi }{d}} \right)} \right)} - \pi \int_{ - \pi /d}^{\pi /d} {\frac{{dk}}{{2\pi /d}}\sum\limits_{n \in occ} {{Q_{2,n,k}}} }\\
& = \sum\limits_{n \in occ} {{\rm Arg}\left( {{\zeta _n}\left( 0 \right){\zeta _n}\left( {\frac{\pi }{d}} \right)} \right)} - \pi {Q_2},
\end{align}
where $Q_{2}$ is the total charge in $\Omega_{2}$ for all the occupied bands.

\section{ii) Derivation of $\sum\limits_{n \in occ} {{{\left| {{\alpha _{n,l}}\left( k \right)} \right|}^2}} = 0.5$ $\forall $ $l,k$}

We consider a nearest-neighbor tight-binding model for gapped GNRs with spatial symmetries. The Bloch wave function is represented as 
\begin{equation}
{\psi _{n,k}}\left( {\bf{r}} \right) = \sum\limits_{l,m} {{\alpha _{n,l}}\left( k \right){e^{ikmd}}{\chi _{m,l}}\left( {\bf{r}} \right)} \label{periodic_gauge}
\end{equation}
in the periodic gauge, where ${{\chi _{m,l}}\left( {\bf{r}} \right)}$ is the atomic orbital $l$ in unit cell $m$. The Hamiltonian can be written as $H = {t_1}\sum\limits_{\left\langle {{{\bf{r}}_i},{{\bf{r}}_j}} \right\rangle } {\left( {c_A^\dag \left( {{{\bf{r}}_i}} \right)c_B^{}\left( {{{\bf{r}}_j}} \right) + h.c.} \right)} $ where $t_{1} \approx -2.7$ eV is the nearest-neighbor hopping parameter~\cite{grapheneTB}, $\left\langle {{{\bf{r}}_i},{{\bf{r}}_j}} \right\rangle$ means that the summation is over the nearest neighbors, and $c_{A}^\dag \left( {{{\bf{r}}_i}} \right)$ is the creation operator for the $p_{z}$ orbital of a carbon atom at position ${\bf r}_{i}$ in the sublattice A. The number of carbon atoms belonging to the A(B) sublattice within one unit cell is $N_A$($N_B$). If we use the Born-von Karman boundary condition ${\psi _{n,k}}\left( {{\bf{r}} + L{\bf{d}}} \right) = {\psi _{n,k}}\left( {\bf{r}} \right)$, where $\bf d$ is the 1D lattice vector and $L$ is the number of unit cells, there will be at least $L\left| {{N_A} - {N_B}} \right|$ zero energy modes because the nearest-neighbor hopping model is chiral-symmetric~\cite{chiral1}. However, since we focus on a system that is gapped, we have $N_{A} = N_{B}$. Furthermore, since the system is chiral-symmetric, the valence bands (VBs) and conduction bands (CBs) must be a mirror image of each other (as shown below). In such a system the half filling corresponds to the occupation of all VBs. We will let $N_A=N_B=N_s$ in the following derivation.

After a Fourier transformation, $H = \sum\limits_k {{\psi _k}^\dag \mathcal{H}\left( k \right){\psi _k}} $, where ${\psi _k}^\dag  = \left[ {\begin{array}{*{20}{c}}
{{\psi _{k,A}}^\dag }&{{\psi _{k,B}}^\dag }
\end{array}} \right]$ with ${{\psi _{k,A}}^\dag }$ (${{\psi _{k,B}}^\dag }$) being the collective creation operator for all carbon atoms belonging to the A(B) sublattice. Namely, ${{\psi _{k,A}}^\dag }$ and ${{\psi _{k,B}}^\dag }$ are both row vectors with a dimension of $1 \times N_{s}$. Since we only consider the nearest-neighbor hopping, the Bloch Hamiltonian $\mathcal{H}\left( k \right)$ can be written as 
\begin{equation}
\mathcal{H}\left( k \right) = \left[ {\begin{array}{*{20}{c}}
0&{{\mathcal{H}_{AB}}\left( k \right)}\\
{{\mathcal{H}_{BA}}\left( k \right)}&0
\end{array}} \right].
\end{equation}
Because $\mathcal{H}\left( k \right)$ is a Hermitian matrix, we have ${\mathcal{H}_{BA}}\left( k \right) = {\mathcal{H}_{AB}}\left( k \right) ^{\dag}$, and ${{\mathcal{H}_{AB}}\left( k \right)}$ is a $N_{s} \times N_{s}$ matrix.

We now define the matrix representation of the chiral operator in this system as 
\begin{equation}
\Gamma  = \left[ {\begin{array}{*{20}{c}}
{{{\left[ 1 \right]}_{N_s \times N_s}}}&0\\
0&{{{\left[ { - 1} \right]}_{N_s \times N_s}}}
\end{array}} \right],
\end{equation}
which means that the matrix elements are $+1$($-1$) for the A(B) sublattice. Since the system has the chiral symmetry, we have 
\begin{equation}
\Gamma \mathcal{H}\left( k \right){\Gamma ^{ - 1}} =  - \mathcal{H}\left( k \right).
\end{equation}
Therefore, the eigenvalue equation $\mathcal{H}\left( k \right)\left[ {{\psi_{n,k}}} \right]  = {E_{n,k}}\left[ {{\psi_{n,k}}} \right] $ leads to
\begin{equation}
\mathcal{H}\left( k \right)\Gamma \left[ {{\psi_{n,k}}} \right]  =  - {E_{n,k}}\Gamma \left[ {{\psi_{n,k}}} \right] ,
\label{chiral}
\end{equation}
which means that (1) if we know all the VB eigenvectors of $\mathcal{H} \left ( {k} \right )$, all the CB eigenvectors can be obtained by operating $\Gamma$ on VB eigenvectors, and (2) the VBs and CBs are a mirror image of each other. Since $\mathcal{H} \left ( {k} \right )$ is a Hermitian matrix, it can be diagonalized by an unitary matrix $U = \left[ {\begin{array}{*{20}{c}}
{\left[ {{\psi_{1,k}}} \right] }& \cdots &{\left[ {{\psi_{2N_s,k}}} \right] }
\end{array}} \right]$. Using the property of the unitary matrix, namely $U{U^\dag } = 1$, we have 
\begin{equation}
\sum\limits_{n = 1}^{2N_s} {{{\left| {{\alpha _{n,l}}\left( k \right)} \right|}^2} = 1,\forall l,k}\ ,
\label{unitary}
\end{equation}
where ${\alpha _{n,l}}\left( k \right)$ is the $l^{th}$ component of the eigenvector $\left[ {{\psi _{n,k}}} \right]$ for ${\mathcal{H}\left( k \right)}$. If we arrange the band index $n$ according to the energy eigenvalues, the VBs and CBs belong to $n=1 \sim N_{s}$ and $N_{s}+1 \sim 2N_{s}$, respectively. If we denote the indexes of carbon atoms belonging to the A(B) sublattice as $l_{A}$($l_{B}$), the chiral symmetry will relate the VB and CB eigenvectors by
\begin{equation}
 \alpha_{n,l_{A}} \left( {k} \right) = \alpha_{2N_{s}+1-n,l_{A}} \left( {k} \right), 
 \label{chiral_A}
\end{equation}
and
\begin{equation}
 \alpha_{n,l_{B}} \left( {k} \right) = -\alpha_{2N_{s}+1-n,l_{B}} \left( {k} \right).
 \label{chiral_B}
\end{equation}
Combining Eqs.~\ref{chiral_A} and~\ref{chiral_B}, we have now proved a useful property for gapped GNRs:
\begin{equation}
\sum\limits_{n \in occ}^{} {{{\left| {{\alpha _{n,l}}\left( k \right)} \right|}^2} = 0.5},\ \forall l,k\ .
\label{chiralresult}
\end{equation}

\section{iii) Derivation of $\gamma'_{2} = \gamma_{2} - M\pi$ and $\gamma'_{1} = \gamma_{1} + M\pi$}
We again let the dimension of ${\mathcal{H}\left( k \right)}$ be $2N_{s} \times 2N_{s}$ and write the tight-binding wave function in the periodic gauge as in Eq.~\ref{periodic_gauge}. The total Zak phase $\gamma$ up to half filling can be calculated as
\begin{equation}
\gamma  = i\sum\limits_{n \in occ} {\int_{ - \pi /d}^{\pi /d} {dk\sum\limits_{j = 1}^{2{N_s}} {\left( {{e^{ik{t_j}}}{\alpha _{n,j}}^*\left( k \right)} \right){\partial _k}\left( {{e^{-ik{t_j}}}{\alpha _{n,j}}\left( k \right)} \right)} } } = \gamma_{1} + \gamma_{2}
\end{equation}
based on the separation scheme in the main text, and $t_{j}$ is the basis vector in the unit cell measured from the real-space origin,
\begin{equation}
{\gamma _1} = \sum\limits_{n \in occ} {\int_{ - \pi /d}^{\pi /d} {dk\sum\limits_{j = 1}^{2{N_s}} {{t_j}{{\left| {{\alpha _{n,j}}\left( k \right)} \right|}^2}} } }, \label{gamma_1_TB}
\end{equation}
and
\begin{equation}
{\gamma _2} = i\sum\limits_{n \in occ} {\int_{ - \pi /d}^{\pi /d} {dk{{\left[ {{\psi _{n,k}}} \right]}^\dag }{\partial _k}\left[ {{\psi _{n,k}}} \right]} } . \label{gamma_2_TB}
\end{equation}
We note that Eq.~\ref{gamma_2_si} summing up to half filling and Eq.~\ref{gamma_2_TB} are equivalent up to a $2\pi$ ambiguity.

For a specific GNR system, we choose 2 different unit cells $\Omega$ and $\Omega'$ and denote their origin-independent part of the Zak phase as $\gamma_{2}$ and $\gamma'_{2}$, respectively. If we change the unit cell from $\Omega$ to $\Omega'$, the eigenvectors of the Bloch Hamiltonian $\mathcal{H} \left( {k} \right)$ and $\mathcal{H}'\left( {k} \right)$ will be related by an unitary transformation
\begin{equation}
\left[ {\psi'_{n,k}} \right]  = U_{k}\left[ {{\psi_{n,k}}} \right] 
\end{equation}
with a diagonal $U_{k} = {e^{i{n_1}kd}} \oplus {e^{i{n_2}kd}} \oplus  \cdots  \oplus {e^{i{n_{2N_{s} - 1}}kd}} \oplus {e^{i{n_{2N_{s}}}kd}}$, where ${n_j}$ is an integer specifying the shuffling of the $j^{th}$ carbon atom measured by the 1D lattice vector. For example, Figs.~1 (a) and (b) in the main text give two different unit cells for the same cove-edged GNR. When transforming from unit cell (a) to unit cell (b), the carbon atoms marked by the green squares are included, which were originally in an adjacent unit cell specified by lattice vector $\bf d$. Therefore they will have $n_{j}=+1$ while the other carbon atoms have $n_{j}=0$. Similarly, when transforming from unit cell (a) to unit cell (c), the additional carbon atoms marked by orange squares were originally in an adjacent unit cell specified by lattice vector $-\bf d$. Therefore, they will have $n_{j}=-1$.

Using Eq.~\ref{gamma_2_TB}, we have the origin-independent part of Zak phase for unit cell $\Omega'$ as
\begin{align}
 {\gamma' _2} &= i\sum\limits_{n \in occ} {\int\limits_{ - \pi /d}^{\pi /d} {dk{{\left[ {{\psi' _{n,k}}} \right]}^\dag }{\partial _k}\left[ {{\psi' _{n,k}}} \right]} }  = i\sum\limits_{n \in occ} {\int\limits_{ - \pi /d}^{\pi /d} {dk{{\left[ {{\psi _{n,k}}} \right]}^\dag }{U_k}^\dag \left( {{\partial _k}{U_k}\left[ {{\psi _{n,k}}} \right]} \right)} }  \\
&  = i\sum\limits_{n \in occ} {\int\limits_{ - \pi /d}^{\pi /d} {dk{{\left[ {{\psi _{n,k}}} \right]}^\dag }{U_k}^\dag \left( {{\partial _k}{U_k}} \right)\left[ {{\psi _{n,k}}} \right]} }  + i\sum\limits_{n \in occ} {\int\limits_{ - \pi /d}^{\pi /d} {dk{{\left[ {{\psi _{n,k}}} \right]}^\dag }{\partial _k}\left[ {{\psi _{n,k}}} \right]} }  \\
&  = i\sum\limits_{n \in occ} {\int\limits_{ - \pi /d}^{\pi /d} {dk{{\left[ {{\psi _{n,k}}} \right]}^\dag }{U_k}^\dag \left( {{\partial _k}{U_k}} \right)\left[ {{\psi _{n,k}}} \right]} }  + {\gamma _2}. \label{eq89}
\end{align}
Since ${{U_{k}}^\dag }\left( {{\partial _k}U_k} \right) = id \left({ {n_1} \oplus {n_2} \oplus  \cdots  \oplus {n_{2N_{s} - 1}} \oplus {n_{2N_{s}}}} \right)$,
the 1$^{st}$ term in Eq.~\ref{eq89} can be simplified:
\begin{align}
& i\sum\limits_{n \in occ} {\int\limits_{ - \pi /d}^{\pi /d} {dk{{\left[ {{\psi _{n,k}}} \right]}^\dag }{U_k}^\dag \left( {{\partial _k}{U_k}} \right)\left[ {{\psi _{n,k}}} \right]} }  \\
&  =  - d\sum\limits_{n \in occ} {\int\limits_{ - \pi /d}^{\pi /d} {dk{{\left[ {{\psi _{n,k}}} \right]}^\dag }\left( {{n_1} \oplus {n_2} \oplus  \cdots  \oplus {n_{2N_{s} - 1}} \oplus {n_{2N_{s}}}} \right)\left[ {{\psi _{n,k}}} \right]} }  \\
&  =  - d\sum\limits_{n \in occ} {\int\limits_{ - \pi /d}^{\pi /d} {dk\left( {\sum\limits_{j = 1}^{{2N_s}} {{n_j}{{\left| {{\alpha _{n,j}}\left( k \right)} \right|}^2}} } \right)} } \label{pass1} \\
&  =  - d\int\limits_{ - \pi /d}^{\pi /d} {dk\left( {\sum\limits_{j = 1}^{{2N_s}} {0.5{n_j}} } \right)}  =  - 2\pi \sum\limits_{j = 1}^{{2N_s}} {0.5{n_j}}  =  - \pi \sum\limits_{j = 1}^{{2N_s}} {{n_j}} . \label{pass2}
\end{align}
Note that we assume that the system has chiral symmetry (Eq.~\ref{chiralresult}) in passing from Eq.~\ref{pass1} to Eq.~\ref{pass2}. Therefore we have
\begin{equation}
{\gamma'_{2}} = \gamma_{2}  - \pi \sum\limits_{j = 1}^{{2N_s}} {{n_j}},\ {n_j} \in \mathbb{Z}.
\label{changeformula}
\end{equation}
In the main text we write $\sum\limits_{j = 1}^{2{N_s}} {{n_j}}  = M \in \mathbb{Z}$ for simplicity. 

For $\gamma_{1}$, we start with Eq.~\ref{gamma_1_TB} as the result for unit cell $\Omega$. As shown above, changing the unit cell from $\Omega$ to $\Omega'$ will shuffle the $j^{th}$ carbon site by an $n_{j}$ lattice vector. Therefore 
\begin{equation}
{\gamma'_{1}} = \sum\limits_{n \in occ} {\int_{ - \pi /d}^{\pi /d} {dk\sum\limits_{j = 1}^{2{N_s}} {\left( {{t_j} + {n_j}d} \right){{\left| {{\alpha _{n,j}}\left( k \right)} \right|}^2}} } } .
\end{equation}
According to Eq.~\ref{chiralresult}, if the system has chiral symmetry, the summation of ${{\left| {{\alpha _{n,j}}\left( k \right)} \right|}^2}$ up to half filling is $0.5$, which is independent of the orbital index $j$ and crystal momentum $k$. It then follows that
\begin{equation}
{\gamma'_{1}} = \gamma_{1}  + \pi \sum\limits_{j = 1}^{2{N_s}} {{n_j}} = \gamma_{1} + M\pi.
\label{gamma1_change}
\end{equation}

\clearpage

\section{iv) Band structures for the cove-edged GNRs}

\begin{figure}
  \includegraphics[scale=0.25]{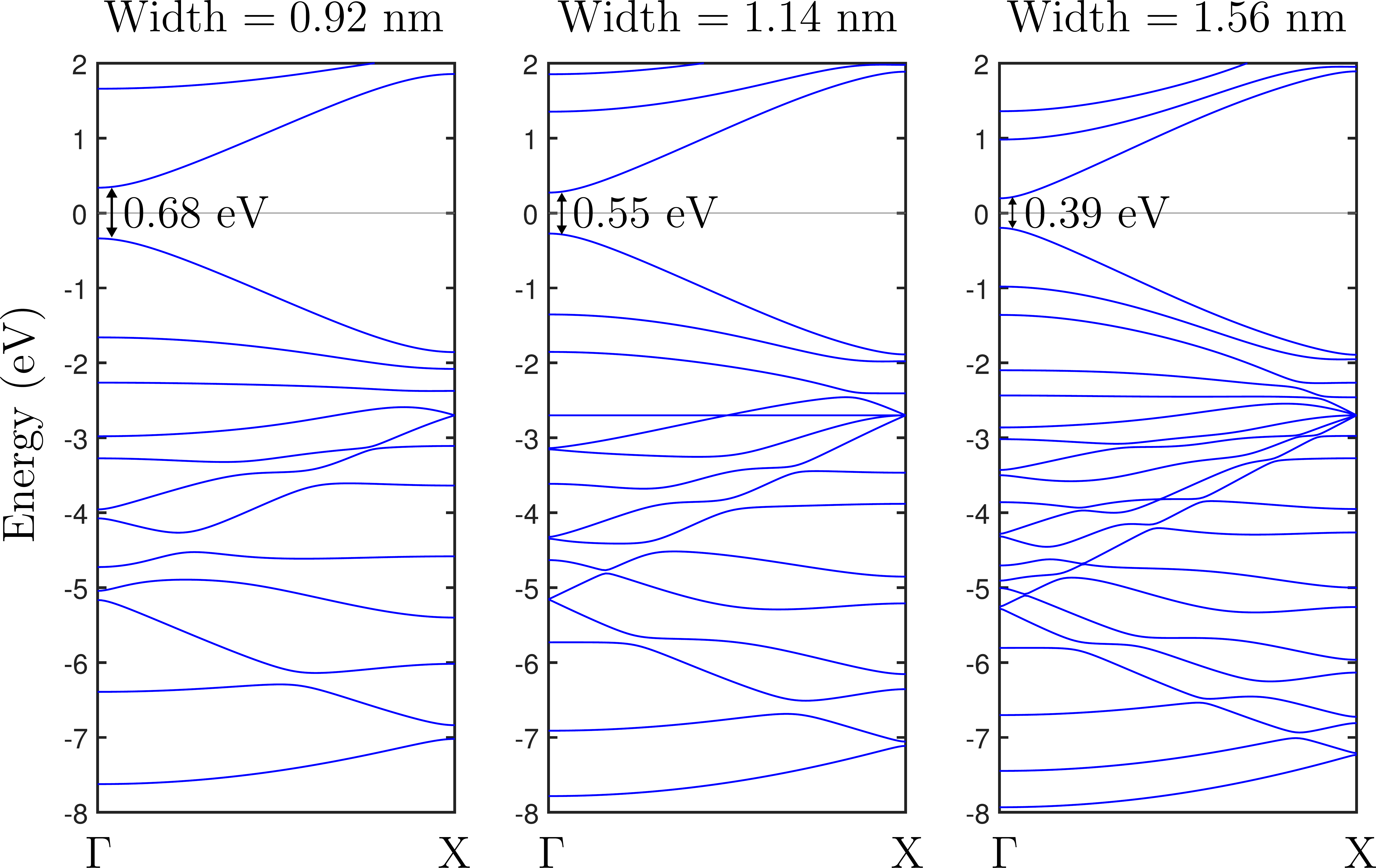}
  \caption{Band structures of cove-edged GNRs with a width of 0.92, 1.14, and 1.56 nm.}
  \label{S2}
\end{figure}

We show the band structures of the cove-edged GNRs considered in this work in Fig.~\ref{S2}. The direct band gaps are indicated. For the cove-edged GNRs with a width of 0.92 nm and 1.56 nm, there is no band crossing at general $k$-points after a zoom-in examination, although some bands do get close. For the coved-edged GNR with a width of 1.14 nm, there is actually one accidental crossing at $k \approx \pi/(2d)$ where $-k \ne  k$ (mod $2 \pi /d$), which is not an inversion invariant $k$-point. A careful examination of the wave functions concludes that we can identify the bands by a continuous gauge on the tight-binding wave functions going through the accidental crossing. Therefore, for these three cases we have examined, the band crossings do not affect our results.

\bibliographystyle{unsrt}
\bibliography{achemso-demo}